# Title

MAPS: Masked Attribution-based Probing of Strategies – A computational framework to align human and model explanations

## Authors

Sabine Muzellec[1*], Yousif Kashef Alghetaa[1,2], Simon Kornblith[3], and Kohitij Kar [1*]

## Affiliation

1. York University, Department of Biology and Centre for Vision Research, Centre for Integrative and Applied Neuroscience, Toronto, Canada
2. University of Toronto
3. Anthropic PBC

   * Correspondence should be addressed to Sabine Muzellec, and Kohitij Kar
E-mail: sabinem@yorku.ca, k0h1t1j@yorku.ca

## Conflict of interest

The author declares no competing financial interests.

## Acknowledgments

KK has been supported by funds from the Canada Research Chair Program (CRC-2021-00326), Google Research, Brain-Canada Foundation (2023-0259), the Canada First Research Excellence Funds (VISTA Program), and the National Sciences and Engineering Research Council of Canada (NSERC, RGPIN-2024-06223). SM is funded by the Connected Minds Postdoctoral Fellowship (supported by CFREF). We thank the members of the ViTA Lab for helpful discussions and comments.

## Data Availability

All data collected for this study and needed to replicate the results can be found at https://osf.io/t6kw8/

## Code Availability

All code needed to reproduce the results can be found at https://github.com/vital-kolab/maps

# Abstract

Human core object recognition depends on the selective use of visual information, but the strategies guiding these choices are difficult to measure directly. We present MAPS (Masked Attribution-based Probing of Strategies), a behaviorally validated computational tool that tests whether explanations derived from artificial neural networks (ANNs) can also explain human vision. MAPS converts attribution maps into explanation-masked images (EMIs) and compares image-by-image human accuracies on these minimal images with limited pixel budgets with accuracies on the full stimuli. MAPS provides a principled way to evaluate and choose among competing ANN interpretability methods. In silico, EMI-based behavioral similarity between models reliably recovers the ground-truth similarity computed from their attribution maps, establishing which explanation methods best capture the model's strategy. When applied to humans and macaques, MAPS identifies ANN-explanation combinations whose explanations align most closely with biological vision, achieving the behavioral validity of Bubble masks while requiring far fewer behavioral trials. Because it needs only access to model attributions and a modest set of behavioral data on the original images, MAPS avoids exhaustive psychophysics while offering a scalable tool for adjudicating explanations and linking human behavior, neural activity, and model decisions under a common standard.

# Introduction

Understanding how humans extract and use visual information to make recognition decisions remains a key goal in vision science[1,2]. When we briefly look at an image—whether it is a dog in the snow or a blurred face—we do not simply use all pixels equally in the decision making; instead, we selectively prioritize certain regions, features, and patterns. These visual strategies, or explanations for our own perceptual judgments, are often implicit and can vary substantially across images depicting the same object category. Identifying these strategies is critical for building explanatory models of perception, for linking behavioral choices to neural computations, and for assessing whether computational models of vision capture not only what humans decide, but how those decisions are formed.

A long tradition in psychophysics has developed tools to probe such strategies indirectly. The *Bubbles* method[3] reveals random, sparse regions of an image and measures how their visibility influences recognition. While powerful for uncovering diagnostic pixels, this approach might arguably[4] sample features in isolation, missing interactions that may be critical for recognizing complex objects or scenes[5]. Although Bubbles offers a useful benchmark, it requires thousands of trials and assumes stable weighting of features across repetitions—assumptions that limit its scalability and ecological validity. Thus, while Bubbles established an important reference, it does not scale as a community resource for benchmarking human and model strategies.

In parallel, the rise of deep artificial neural networks (ANNs) has created new opportunities to model human vision[6]. Modern ANNs achieve human-level accuracy on many object recognition benchmarks[6,7], and their internal representations show partial alignment with primate ventral visual stream activity[8,9]. This has led to a natural question: can we use ANNs as computational microscopes to study human perceptual strategies? We reason that if an ANN uses similar features to a human for the same task, its internal explanation for a decision might serve as a proxy for the human explanation. In machine learning, such explanations are commonly generated by explainable AI (XAI) techniques—such as saliency maps[10], deconvolution[11], guided backpropagation[12], and integrated gradients[13]—that assign an importance score to each pixel for a given decision. These methods produce visually compelling heatmaps (also referred to as attribution maps) that resemble, in spirit, human feature maps from psychophysics.

However, applying such attribution methods directly to humans is not feasible. Many of these techniques depend on differentiating a model's internal parameters with respect to its inputs—computing gradients or perturbations within a fully specified network. Human visual systems, by contrast, do not provide such access: we cannot derive a brain-wide importance map by mathematically differentiating neural responses to specific image pixels. Even advanced neuroimaging methods offer only indirect correlates of population activity and cannot isolate the contribution of each visual feature to a perceptual decision. Moreover, even when attribution maps can be precisely computed in artificial networks, they vary widely across explainable AI (XAI) algorithms[14,15]. The same network can yield qualitatively different saliency structures depending on the chosen attribution rule, demonstrating that explanation choice is itself nontrivial[16]. This highlights a fundamental measurement gap—we currently lack a principled, behaviorally grounded framework to determine which explanations meaningfully correspond to human visual strategies.

Here, we introduce MAPS (Masked Attribution-based Probing of Strategies), a behaviorally validated computational tool designed to close this gap. MAPS converts attribution maps into explanation-masked images (EMIs)—minimal images restricted to a low pixel budget—and asks whether they reproduce the *image-by-image pattern of recognition performance* observed with the original, clean images. A key innovation of MAPS is that it provides a principled way to evaluate and choose among competing ANNs and interpretability methods. By comparing EMI-driven behavioral similarity between models to the ground-truth similarity derived from full attribution maps, MAPS identifies which explanations best capture model strategy. Applied across humans, macaques, and neural data, MAPS functions as a unifying benchmark to identify ANN–explanation combinations that best align with biological vision.

Our contribution is not a new XAI algorithm, but a general, experimentally tractable framework for *measuring* alignment between human visual strategies and model explanations. This addresses a gap left by both traditional human-probing methods and contemporary XAI evaluations. Whereas methods like Bubbles[3] and classification images[17] can be slow, noisy, and hard to compare to ANN outputs, MAPS offer a direct, scalable way to test whether masking out a model's most important pixels has the same behavioral consequences for humans. Because the method separates a calibration step (ANN–ANN validation) from a deployment step (ANN–human comparison), it can be standardized across various different behavioral tasks, enabling cumulative benchmarking of human–model strategy alignment.

The implications of MAPS extend beyond human–ANN comparisons. Identifying ANN + XAI combinations that replicate human behavioral effects yields concrete hypotheses about the neural mechanisms of primate vision, testable with neural data. The same logic applies to other domains—audition[18], language[19], or motor control[20]—where internal access is limited but behavioral perturbations can be engineered. In all cases, MAPS focuses on the information humans actually use to decide, rather than on visualizing model attributions. In summary, MAPS offers a principled, scalable, behaviorally grounded framework for probing and benchmarking human–model alignment, closing a key methodological gap between perception, computation, and neural implementation.

# Results

To understand whether ANNs and biological object vision rely on comparable visual strategies we first need a common behavioral foundation. We therefore began by developing a well-controlled match to sample binary object discrimination task and established corresponding benchmarks to measure how humans use visual information under time-limited conditions. This initial behavioral analysis provides a reference for what constitutes effective visual explanations—how selective sampling of image features supports recognition performance. To ground this framework empirically, we first examined how human observers perform the task itself. By quantifying both accuracy and the diagnostic features that guide decisions, we sought to reveal the visual strategies underlying human object discrimination—providing a baseline against which model-derived explanations can later be evaluated.

## Human object discrimination depends on selective visual strategies

We began by quantifying human performance on a rapid match-to-sample object recognition task with ten object categories, and 20 images per category (**Figure 1A**). On each trial, participants viewed a sample *Test* image for 100 ms (consistent with the concept of core object recognition[2]) followed by two canonical choice tokens (containing the target object and one distractor object) and reported which one matched the sample. Although accuracy measures overall task performance, the critical question is which visual information participants rely on to make their decisions. Multiple strategies are possible: observers may base their responses on the object region itself (**Figure 1B**, left panel: object only), or they may use more selective diagnostic features distributed across the image (**Figure 1B**, right panel: Bubbles based). To probe these possibilities, we compared two established methods (on a subset of 40 images): (i) removing the background and isolating the object (object-only), and (ii) applying Bubble masks[3], which reveal sparse subsets of pixels across trials.

In both cases, the expectation is that accuracy and consistency with behavior on the original, unaltered (henceforth referred to as clean) images should stay high over a range of reduced visible-pixel budgets before decreasing, reflecting reliance on selective diagnostic features rather than indiscriminate use of all available pixels.

Empirically, we observed this predicted dissociation. Object-only masks supported high accuracy even with very few pixels (**Figure 1C** – purple dot, number of pixels = 90.88% (mean) ±1.25(s.e.m), accuracy = 0.93 (mean) ±0.09 (s.e.m)), but failed to predict the pattern of decisions observed with clean images (**Figure 1D** – purple dot, consistency = 0.38 (mean) ±0.04 (s.e.m)). By contrast, Bubble-based explanations showed an initial plateau followed by a declining accuracy as pixel budgets decreased (**Figure 1C** – red, accuracy = 0.83 (mean) ±0.12 (s.e.m) at 80% pixel budget), and more accurately reproduced clean-image behavioral patterns (**Figure 1D** – red, consistency = 0.56 (mean) ±0.04(s.e.m) at 80% pixel budget). Thus, Bubble methods provided a better account of the visual strategies guiding human recognition.

However, Bubble methods are labor-intensive, as each data point reflects the aggregation of many randomized trials per image. For the present experiment, we generated Bubble stimuli by overlaying random Gaussian transparency masks on a set of 40 base images drawn from a set of 200 images. For each image, 50 unique trials were created, each containing 20 circular Gaussian bubbles ($\sigma$ = 20 pixels) randomly positioned across the image, revealing only small subsets of diagnostic pixels against a uniform gray background. This yielded 2,000 masked stimuli per session, with 18 repetitions per image and a total of approximately 36,000 human behavioral trials. To combine results across trials, we reconstructed composite Bubble maps by weighting each mask according to behavioral accuracy and then blending the most informative regions back into the original image. These composite maps provided per-image predictions of the visual information most used during the object discrimination tasks by human participants. To systematically evaluate how recognition performance varied with the proportion of visible pixels, we generated and tested seven distinct Bubble-derived percentile cutoffs—5%, 10%, 20%, 25%, 30%, 40%, and 50% of the most diagnostic regions identified in the composite maps. Each percentile condition was tested in a separate behavioral run, yielding an additional ~10,000 trials in total. This expanded dataset enabled us to quantify both the accuracy and consistency of human object recognition across pixel-visibility levels, establishing a precise behavioral signature of selective feature use that would later serve as the ground truth for evaluating model-based explanations.

This extensive data collection to converge on stable explanations, makes Bubbles impractical for large-scale studies. And this limitation motivates the search for alternative approaches to recover human visual strategies more efficiently. We therefore next asked whether current ANN models of human object recognition could serve as a scalable tool for generating explanations that align with human strategies.

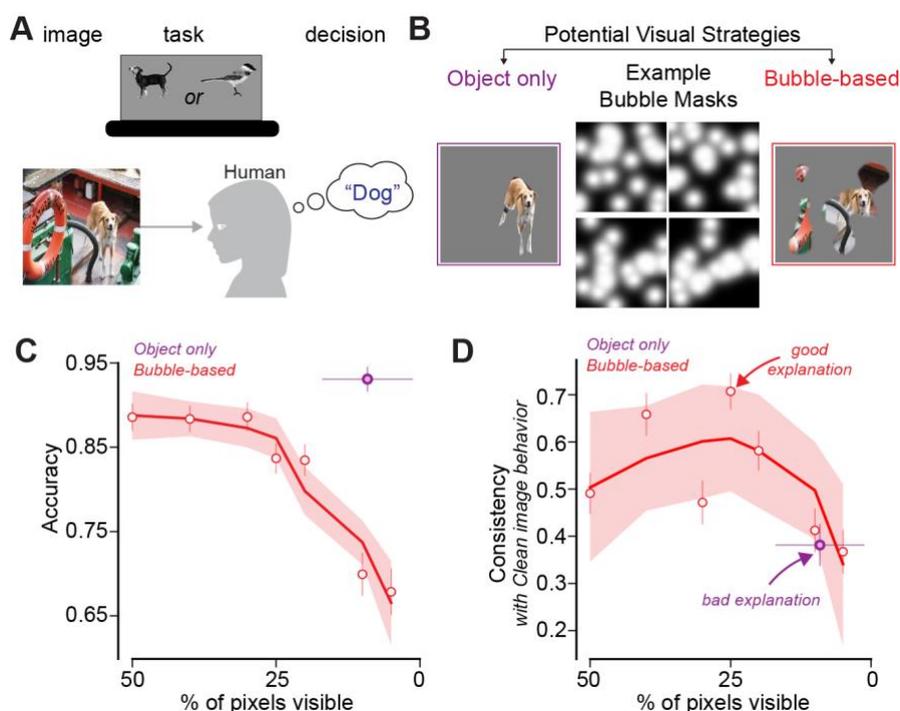

**Figure 1. Benchmarking human visual strategies: Bubbles as a baseline resource. A.** When humans perform a rapid object discrimination task (dog vs. bird), their decisions rely on *visual strategies*—some pixels are more informative than others. In our study, humans performed a match to sample, binary object discrimination task across 10 object categories and 200 images. In brief, on each trial, participants viewed a sample *Test* image for 100 ms, followed by two canonical choice tokens (containing the target object and one distractor object) and reported which one matched the sample. **B.** Candidate strategies could include using only the object region (purple) or feature subsets revealed by Bubble masks (red). **C.** Accuracy as a function of the proportion of visible pixels. Object-only masks (purple) yield consistently high accuracy even with few pixels, whereas Bubble-based masks (red) show decreasing accuracy as fewer pixels are revealed. Errorbars indicate s.e.m across images. **D.** Consistency in image-level accuracy patterns between masked-image performance and clean-image based behavior. Despite reduced accuracy at low pixel budgets, Bubble-based explanations (red) align more closely with the behavioral pattern of clean images ("good explanation"), whereas object-only masks (purple) achieve high accuracy but fail to capture the strategy underlying human performance ("bad explanation"). These results illustrate that human visual strategies are not explained solely by object presence but are better captured by attribution-based probes. Error bars represent 95% confidence intervals, estimated by bootstrapping the image set prior to computing the image-level correlations.

## ANN behavior aligns with humans and enables explanations

ANNs provide a powerful framework for generating hypotheses about how the human visual system performs object recognition[6]. Like humans, ANNs can be evaluated on the same task, yielding behavioral outputs that align closely with human performance[7,21]. In addition, ANNs provide access to their internal unit responses and to explanation maps (**Figure 2A**), which highlight the visual features most influential for a given decision. These explanations offer a direct window into a model's visual strategies—something that cannot be obtained from humans, where only behavioral responses are observable.

Consistent with prior reports[7], we first observed that ANNs achieved high accuracies and image-level consistencies with human behavior on our imageset (e.g., ConvNext[22]: mean accuracy: 0.91, mean consistency: 0.53, number of images: 1280, 128 images per object, 10 objects,40 images used for the held-out psychophysics experiments were not used for fair cross-validation). We also observed that across diverse architectures, model accuracy was strongly correlated with the models' consistency with human behavior at the image-by-image level (Spearman Correlation, *r(9) = 0.90, p < 0.001*; **Figure 2B** and **Figure S1** for a similar result on the ImageNet dataset).  These results indicate that, currently more accurate models not only perform better overall but also tend to make similar errors on  individual images, suggesting a common overall visual strategy (despite evidence of certain clear deviations in their behaviors[23–25]).

To better understand *why* different models, align with human behavior to varying degrees, we next examined *how* they process visual information. In computer vision, a range of explainable AI (XAI) techniques—also referred to as attribution methods—have been developed to probe which parts of an image contribute most to a model's decision (**Figure 2C**, see also **Figure S1** for more examples). Gradient-based methods compute the derivative of the model's output with respect to the input image, capturing how sensitive the prediction is to changes in individual pixels. The simplest example is the saliency map[10], which uses the raw gradient directly. Because raw gradients can be noisy[26], variants such as Input×Gradient[27] and Guided Backpropagation[12] have been introduced to smooth or sharpen the resulting maps, yielding attributions that are often easier to interpret. In contrast, perturbation-based methods[11,28] assess the impact of systematically altering or removing information: occlusion removes local patches of pixels, feature ablation disables internal feature channels, and feature permutation shuffles feature activations. These approaches directly test how much the model's decision depends on particular pixels or features. Although both families aim to highlight diagnostic regions, they often produce qualitatively different maps, underscoring substantial method-dependent variability. Importantly, most of these techniques require complete access to the model's architecture and parameters—something that is not possible in humans.

To systematically quantify variability in explanations, we evaluated a broad set of ANN models and XAI methods. Specifically, we considered 11 ANN architectures (see Methods: *Model Selection*), fine-tuned each on our task (see Methods: *Training Procedure*), and generated explanations using 12 distinct attribution techniques (see Methods: *Generation of Explanations*).

Upon visual inspection (and as suggested in prior work[14]), we observed that the attribution maps from the same model for the same images (e.g. ResNet-50) varied across explanation methods. Therefore, we sought to determine whether these qualitative differences were statistically reliable. To this end, we quantified how strongly explanation variability arises from the XAI method choice versus stochastic variation in network training (**Figure 2D**; see Methods: *Quantifying the variance between methods*). For each test image, we normalized the attribution maps and computed pairwise L2 distances. We then grouped these distances into two complementary conditions: (i) a *between-method* condition, in which we compared explanations generated by different attribution techniques applied to the same network initialization, and (ii) a *between-initialization* condition, in which we compared explanations generated by the same attribution method across independently trained networks. This analysis revealed that attribution maps varied far more across methods than across initializations (Wilcoxon signed-rank one-tailed test: *Z = 0, p < 0.001*), demonstrating that the choice of explanation method—rather than stochastic variation in training—drives most of the variability in model strategies.

Together, these results highlight both the promise and the challenge of using ANNs to study human vision. On one hand, models aligned with human behavior provide scalable access to explanation methods that can reveal candidate visual strategies. On the other hand, explanation methods are highly inconsistent with one another, and equivalent human explanations cannot be generated using the same attribution techniques. Thus, to meaningfully compare models and humans, we must establish a principled approach to identify which model–method combinations yield explanations most aligned with human strategies—without requiring the direct generation of attribution maps for human observers.

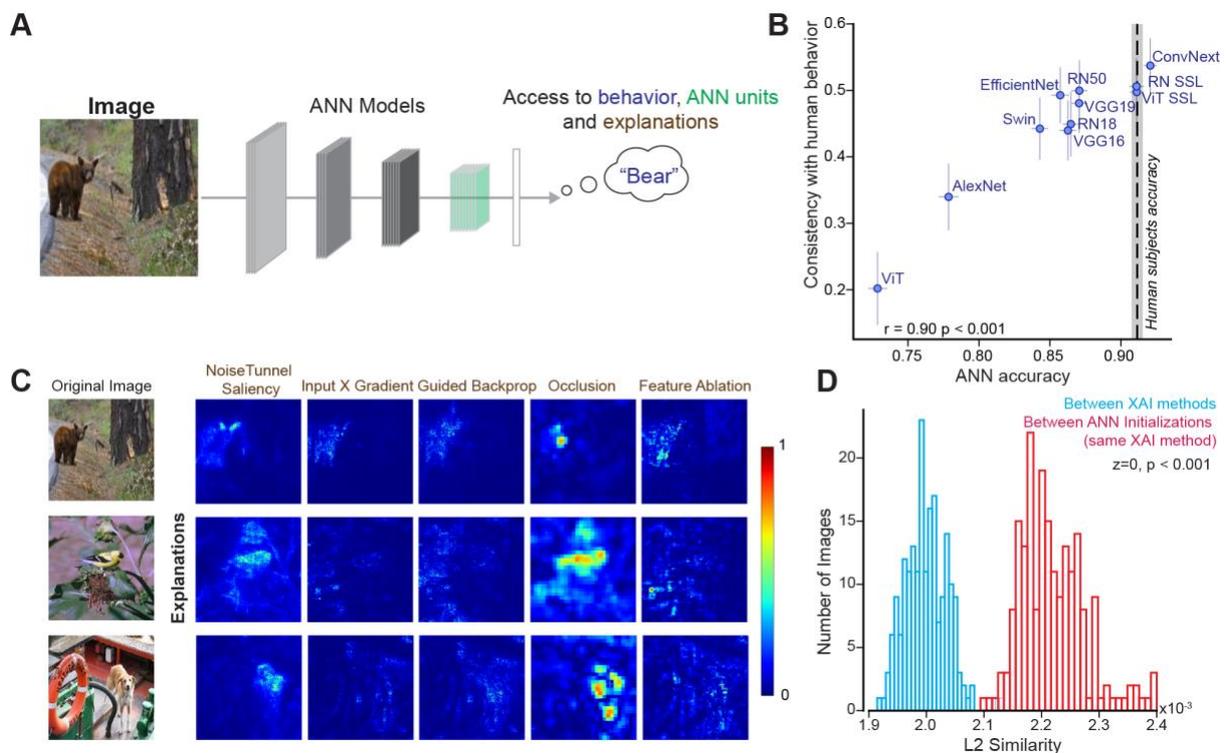

**Figure 2. Establishing ANN behavior as a scalable source of explanations. A.** Like humans, ANN models can be evaluated on the same object discrimination tasks, with access to behavior, unit responses, and explanations. **B.** Relationship between model accuracy and consistency with human behavior. ANN accuracy correlates strongly with human recognition patterns (humans: r = 0.90, p < 0.001), suggesting that explanations derived from high-performing models may approximate human recognition strategies. **C.** Examples of attribution maps for the same image across different XAI methods. NoiseTunnel Saliency, GradientInput, guided backpropagation, occlusion, and feature

ablation highlight distinct regions, underscoring method-dependent variability. **D.** Quantification of variance across explanations. The distribution of L2 similarity values shows that the variance across XAI methods (blue) is significantly larger than the variance across independent ANN initializations (red, $p < 0.001$), indicating that explanation choice dominates over stochastic training differences.

## Estimating the true differences in explanations

The variability across explanation methods (**Figure 2D**) raises a fundamental challenge: how can we determine which explanations best capture visual strategies that generalize across models and, ultimately, to humans? To address this, we developed a framework called *Masked Attribution-based Probing of Strategies (MAPS)*. MAPS is designed to compare explanations without requiring direct access to the internal components of the ANNs (i.e. the neural substrate).

First, we sought to establish a ground-truth measure of similarity between model explanations. For this purpose, we distinguished between two types of systems. A **Reference** is a system to which we have full access—including its internal responses and attribution maps—allowing us to compute explanation similarity directly. A **Target**, in contrast, is a system where we only observe behavior and cannot directly access its explanations. In the present analyses, both Reference and Target were ANNs, but later in the paper we apply the same logic to biological systems such as humans and monkeys.

The first step is to define the Reference model for each Target. To ensure that comparisons are made between behaviorally aligned systems, we chose as the Reference the ANN whose image-level behavioral accuracy patterns were most correlated with those of the Target (**Figure 3A**; see Methods: *Target–Reference pairs* and **Figure S2**). To identify the best model, we performed multiple cross-validated splits, selecting 80 "train" images (from the full test set of 200 MS COCO images, while 40 images were held-out for the psychophysics validation reported below, see Methods: *Cross-Validation*), and selected the model with the highest average correlation on these images for ResNet50[29] as Target, the best reference is VGG19[30] with *r(8) = 0.74, p < 0.001*). To formalize these comparisons, we computed pairwise distances between attribution maps using the L2 norm. The inverse of this distance provided a straightforward similarity score, yielding a natural ranking of explanation closeness across models and images (see Methods: *Ground Truth*). To test the robustness of this choice, we compared L2-based similarity scores with alternatives: L1 distance and the perceptually motivated LPIPS metric. In practice, we randomly selected 50 images, generated the explanations using one model and one method (here ConvNeXt and Noise Tunnel Saliency) and measured pairwise similarities using each of the three measures. As shown in **Figure 3B** (and **Figure S1C**), the three metrics were strongly correlated (L2 vs. L1: *r(48) = 0.68, p < 0.001*, L2 vs. LPIPS: *r(48) = 0.51, p < 0.001*) and produced nearly identical rank orders of similarity. This confirms that the ranking reflects genuine differences in attribution maps rather than artifacts of a particular distance function. Consequently, in the subsequent sections, we use the inverse of the L2 distance as measure quantifying the similarity between explanations.

Once these Reference–Target pairs and the similarity metric were established, the problem reduces to identifying, within each pair, which explanation method provides the most meaningful account of their shared strategy.

We then generated attribution maps for pairs of Reference–Target models on 160 test images and compared them using a given XAI method (**Figure 3C**). Intuitively, if an explanation method produces highly similar attribution maps across the two models, it suggests that the method captures a visual strategy that generalizes from the Reference to the Target. Conversely, divergence between maps indicates that the models rely on distinct visual evidence, and the explanation is less transferable. To ensure generality, we performed this analysis across the 11 different ANN architectures and 12 distinct attribution methods.

This ground-truth ranking provides the foundation for MAPS. It defines how close or far apart different explanations truly are when full access to models is available. In the rest of the analysis, we ask whether this same ranking can be recovered using a behavioral surrogate—placing ourselves in the realistic setting where the Target is treated as a black box and only its behavior is observable.

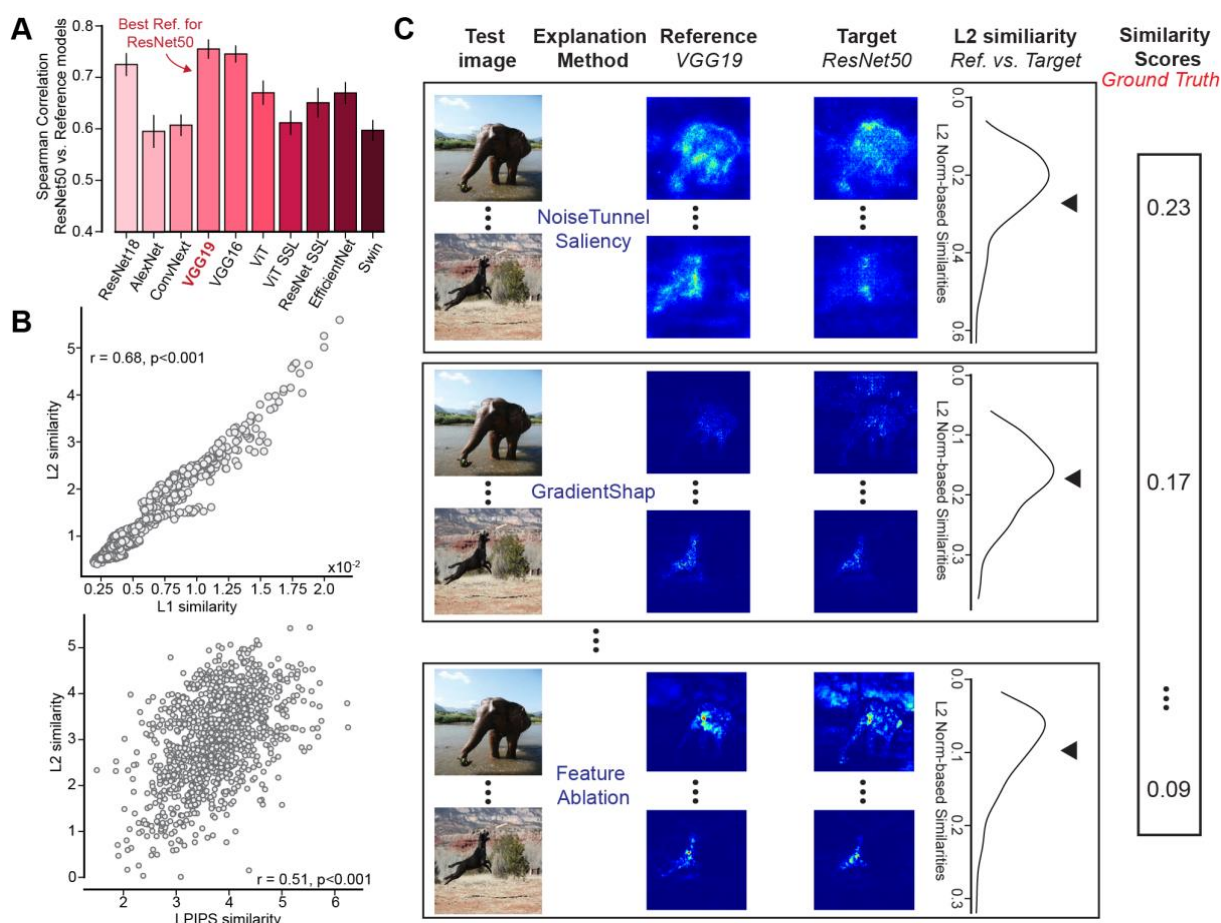

**Figure 3. Ground-truth similarity framework for explanation methods. A.** Behavioral alignment used to define Reference–Target pairs (here ResNet-50 as Target). Bars show the Spearman correlation between ResNet-50 and all candidate Reference models (mean ± s.t.d. across cross-validated image splits). **B.** Correlations between similarity metrics. Pairwise comparisons show that L2, L1, and LPIPS distances are strongly correlated across images, confirming that the rank order of explanation similarity is robust to the choice of distance function. **C.** Example attribution maps for the same test image generated from two models (reference: VGG19; target: ResNet50) using different explanation methods. For each method, the similarity between the reference and target attribution maps was quantified using the L2 norm, yielding similarity scores that provide a ground-truth ranking of explanation closeness.

## Generating explanation-masked images (EMIs)

The next step in building the MAPS framework is to translate explanation similarity into measurable behavioral effects. To do this, we leveraged *Explanation-Masked Images (EMIs)*—filtered versions of the original images that selectively retain pixels based on their attribution scores. EMIs are generated in two steps (**Figure 4A**). First, an attribution map is computed from the Reference model using a chosen explanation method (e.g., saliency, occlusion), which assigns an importance score to each pixel. Second, a percentile cutoff is applied to these scores to produce *EMIs*, in which only the top-ranked pixels are retained. By varying this cutoff, we can systematically test the impact of including progressively more or fewer informative features.

This procedure yields predictable and interpretable effects on model behavior that mirror the Bubbles benchmark. Because the Reference's attribution map identifies which pixels matter for its decisions, progressively removing uninformative pixels while retaining informative ones preserves high recognition performance over a broad range of cutoffs (**Figure 4B**). Accordingly, many model–method pairs exhibit an extended plateau in accuracy: as long as the mask spares the informative pixels, performance remains largely unchanged. Only when the cutoff begins to exclude those informative regions does accuracy drop sharply, producing the characteristic decline at stricter cutoffs. Across architectures, this plateau-then-decline pattern confirms that EMIs modulate behavior in a way that reflects the explanatory content of the Reference's attribution maps, rather than generic pixel loss.

Within the MAPS framework, EMIs serve as the critical link between attribution maps and behavior. By systematically modulating recognition performance, they provide a way to translate explanatory content into behavioral effects. In the next section, we test whether these EMI-driven behaviors can be used to recover the similarity structure between explanations established in our ground-truth analysis.

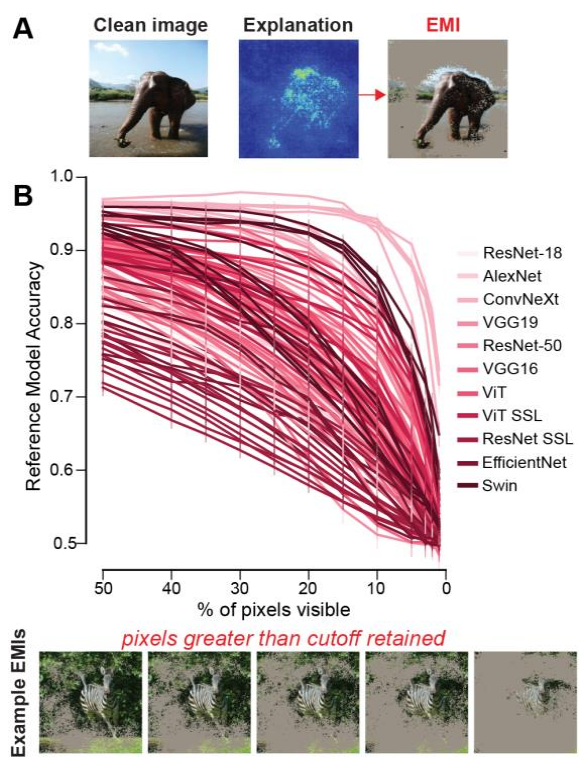

**Figure 4. MAPS pipeline: transforming attribution maps into behavioral probes (EMIs). A.** Example of EMIs, created by retaining only the top-ranked pixels from an attribution map. **B.** Model accuracy on EMIs decreases monotonically as more informative pixels are removed, with the slope varying across architectures. Insets show example EMIs at different percentile cutoffs.

**Validating MAPS: EMIs as a behavioral proxy for explanation similarity**

We next tested whether behavioral similarity on EMIs could serve as a proxy for explanation similarity. The key idea is that if a Reference and a Target model rely on similar visual strategies, they should also produce similar attribution maps. EMIs derived from the Reference would then modulate both systems in a comparable way, such that the image-by-image accuracies of the Target on these EMIs correlate with those of the Reference. In this way, behavioral similarity provides an indirect measure that parallels the ground-truth similarity between their explanations (**Figure 5A**).

To operationalize this, we generated EMIs from the Reference model's attribution maps and presented these identical stimuli to both the Reference and the Target. For every model, attribution method, and EMI set, we then computed image-level behavioral patterns (see Methods: *Proxy based on behavioral similarity across models*). Proxy similarity was quantified as the Spearman correlation between the Reference and Target image-level behaviors, yielding an image-wise proxy profile across percentiles and methods.

Finally, to evaluate whether the EMI proxy captured explanation similarity, we compared the proxy profiles against the ground-truth profiles derived from attribution-map distances (see Methods: *Comparing Proxy and Ground Truth*) on multiple cross-validated splits of the 160 images (using the 80 "test" images different from the ones used to select the best model). For each Reference–Target pair, we asked whether attribution methods that yielded more similar maps also produced stronger behavioral correlations. This prediction yields two sharply contrasting outcomes (illustrated in **Figure S2B)**. If EMI-based correlations are unrelated to the ground-truth distances between explanations, then EMIs fail as a proxy, indicating that behavior alone cannot recover explanation similarity. Conversely, if EMI-based correlations reproduce the rank order established by the ground truth, then EMIs provide a valid behavioral surrogate for comparing explanations.

Across models and explanation methods, this is exactly what we observed. EMI-driven correlations recovered the ground-truth ordering of explanation similarity in most models (**Figure 5B**, Resnet50: *r(11) = 0.70, p = 0.04*). In other words, methods that brought attribution maps closer together also produced stronger alignment in behavior when probed with EMIs. This finding is non-trivial: it shows that recognition performance on filtered stimuli contains enough information to reconstruct the fine-grained similarity structure of explanations, thereby revealing whether two systems rely on comparable visual strategies.

Having established that the EMI-based proxy reliably reflects the similarity of explanations between models, we next asked which attribution methods, on average, yield the most consistent behavioral patterns across models. To address this, we averaged the proxy correlations across all Reference–Target pairs and across multiple cross-validated image splits. This analysis revealed a rank order of methods (**Figure 5C**): Noise Tunnel Saliency produced the strongest behavioral alignment (*r(78) = 0.73, p < 0.001*), while Deconvolution consistently produced the weakest (*r(78) = 0.66, p < 0.001*). Across models and percentiles, we find that Noise Tunnel Saliency consistently outperforms Deconvolution at approximating the model's behavior on clean images (**Figure S3**, Wilcoxon paired t-test: *Z=8.0, p < 0.001*). Thus, MAPS provides a principled way to adjudicate among explanation methods, identifying which techniques most effectively capture shared visual strategies across systems.

Together, these findings provide a critical validation of the MAPS framework. EMIs successfully transform attribution maps into behavioral probes that recover a comparable similarity structure as direct map comparisons, while also enabling systematic selection of the most human-aligned explanation methods. With MAPS validated on models, we next extended the framework to Targets where internal access is almost impossible—such as humans—thereby using artificial networks not just to compare but to *recover* the visual strategies underlying biological recognition.

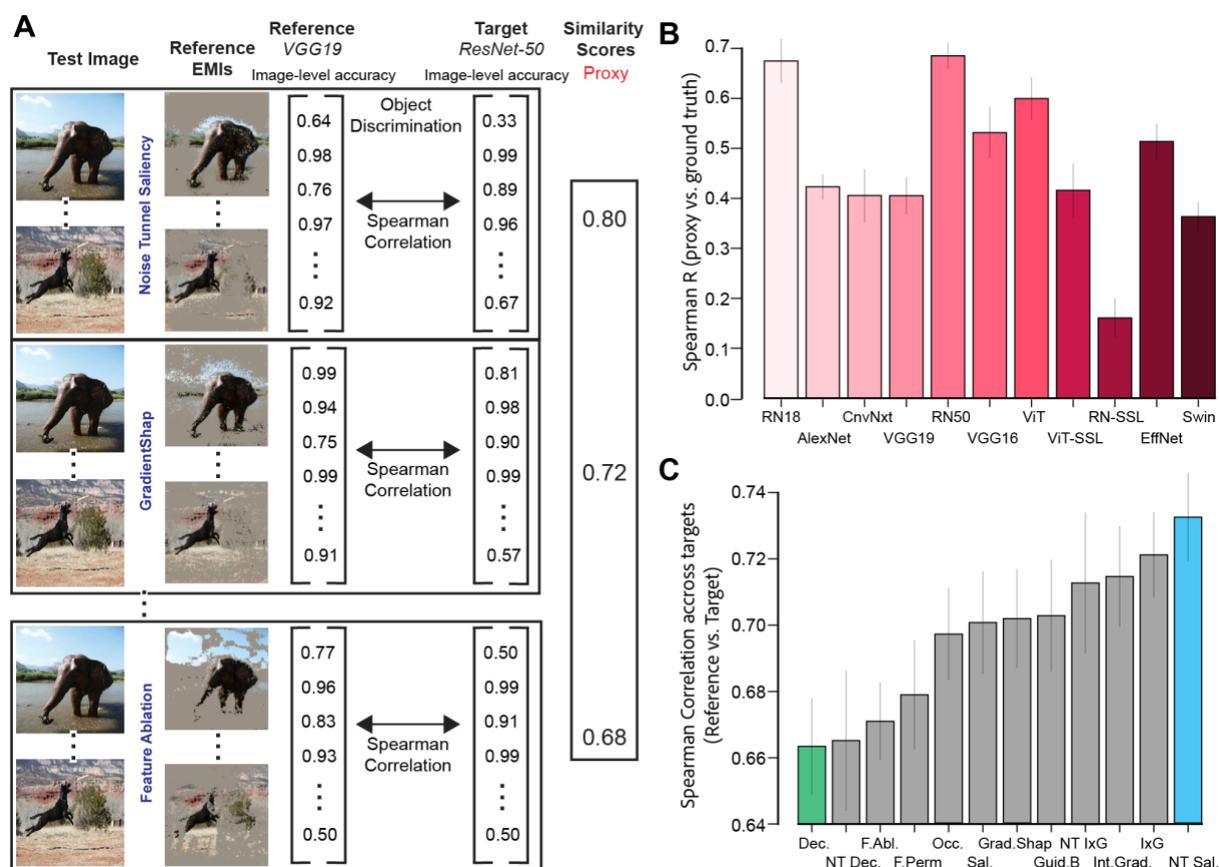

**Figure 5. Validation of EMIs as a resource for recovering explanation similarity. A.** Example of proxy computation. EMIs derived from a reference model (e.g., VGG19) are evaluated on both the reference and a target model (e.g., ResNet-50). Image-level accuracies are collected and correlated across images, yielding a proxy similarity score (e.g., Spearman correlation = 0.80). **B.** Across Reference–Target model pairs, EMI-based proxy correlations vs. ground-truth similarity derived from attribution maps. Bars show the maximal Spearman correlation between proxy and ground truth across percentiles, averaged over multiple cross-validated image splits (error bars: median absolute deviation across splits). **C.** Rank ordering of explanation methods based on their average behavioral correlation across Reference–Target pairs. Bars represent the mean Spearman correlation across cross-validated splits (error bars: median absolute deviation). Full names of the methods from left to right: Deconvolution, Noise Tunnel Deconvolution, Feature Ablation, Feature Permutation, Occlusion, Saliency, Gradient Shap, Guided Backpropagation, Noise Tunnel Input x Gradient, Integrated Gradients, Input x Gradient, Noise Tunnel Saliency.

## Uncovering human visual strategies using MAPS

Having validated MAPS in silico, we next applied the framework to humans as Targets. We tested human participants (n=56) on the binary object discrimination task (**Figure 6A**) using EMIs generated from ConvNext (model best matching human behavior on clean images, see **Figure 2B,** also see Brain-Score[21]). We generated these EMIs using Noise Tunnel Saliency (best method on average, **Figure 5C**) on the set of the held out 40 images (also used to test human subjects on the Bubbles methods and held out from all previous analyses involving ANNs). For comparison, we also generated a separate image-set of EMIs using Deconvolution (worst method on average, **Figure 5C**).

We then compared EMI-based explanations to object-only and Bubble-based stimuli (**Figure 6B**, see Methods: *Object-only stimulus generation*; *Bubbles stimulus generation*). As shown in **Figure 6C-D**, the explanatory value of each method depends strongly on the proportion of visible pixels. Object-only (purple dot) masks maintain high accuracy across pixel budgets but fail to capture consistency with clean-image behavior, confirming that object presence alone is a poor explanation of human recognition. Bubble-based masks (red lines) show the opposite trade-off: as pixel visibility decreases, overall accuracy declines, yet at certain intermediate levels they preserve both relatively high accuracy and high consistency, making them a good probe of human strategies. Strikingly, EMI-based masks (blue lines) derived from the selected model and method pairs recover this same Bubble-like regime. By focusing on the pixels identified by the best ANN–XAI explanations, EMIs reveal a subset of visual features that supports both high accuracy (minimum accuracy = *0.84 ± 0.02* at 5% pixels visible) and high consistency (minimum consistency = *0.40 ± 0.05* at 10% pixels visible) — outperforming the Bubble-like effect (Wilcoxon paired test on accuracies across percentiles: *z=0.0, p=0.016*, and one-sided t-test on consistency values: *t(5) = 1.746, p=0.07*) and capturing the visual strategies underlying human recognition, while requiring far fewer behavioral trials.

Having established that EMIs can recover Bubble-like effects when directly compared with human behavior, we next asked whether MAPS could provide a scalable solution that avoids collecting new human data for every experiment. The critical test was whether model behavior on EMIs could serve as a reliable stand-in for human EMI performance. We therefore compared the image-by-image recognition patterns of humans on clean images to the corresponding patterns predicted by models on EMIs generated from their own explanations (**Figure 6F**). Remarkably, ConvNext with Noise Tunnel Saliency (best method) produced a similar correspondence with human behavior on clean images compared to the Bubbles method (paired t-test: *t(5)=0.432, p=0.684*). In addition, the behavioral alignment of ConvNext with Noise Tunnel Saliency (best method) was significantly better than that using ConvNext with Deconvolution (worst method, see **Figure S5**, paired t-test: *t(5)=1.888, p=0.009*).

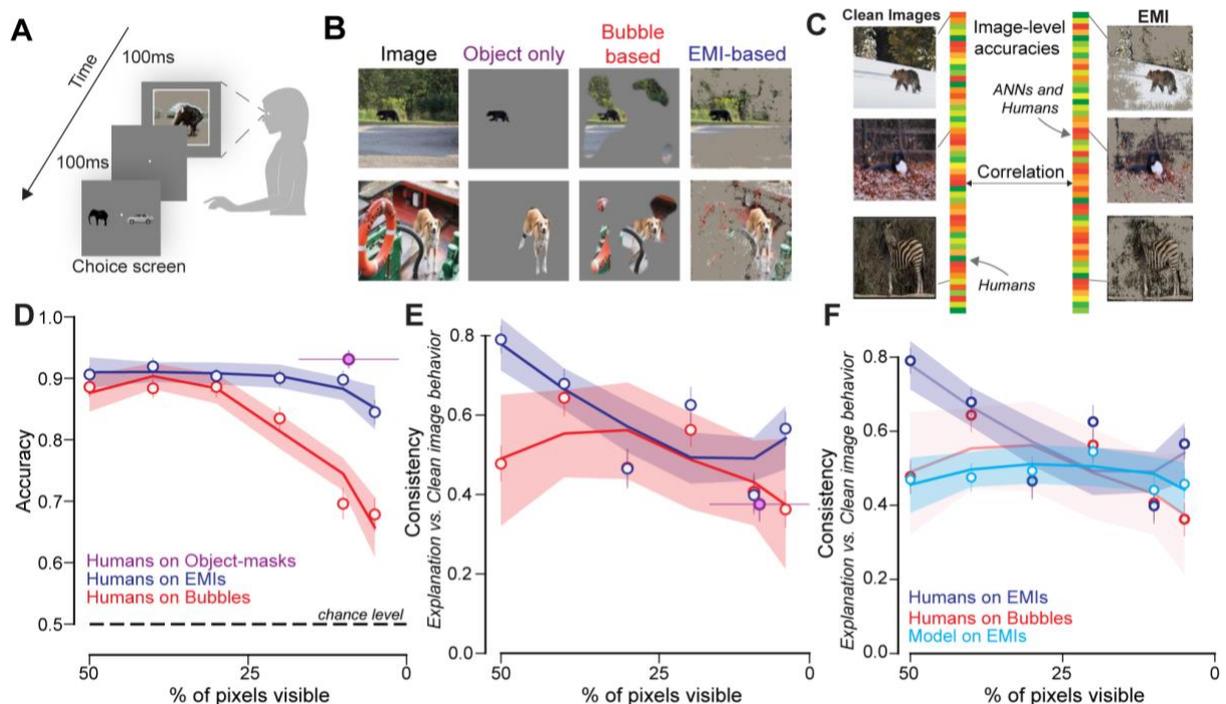

**Figure 6. Human–model alignment benchmarked with MAPS. A.** Human participants performed a rapid object discrimination task on the ANN+XAI based EMIs (100 ms presentation, 100 ms delay) before making a choice. **B.** Example masked stimuli illustrating object-only, Bubble-based, and EMI-based strategies. **C.** Illustration of how image-level accuracies from humans and models were compared. Correlations were computed between human accuracies on clean images and ANN and human accuracies on EMIs generated from the same stimuli. **D.** Human recognition accuracy as a function of visible-pixel percentage. Object-only masks (purple dot, same as Figure 1C) maintain high accuracy independent of pixel budget, whereas Bubbles (red, same as Figure 1C) and EMI-based masks (blue) show the characteristic plateau followed by a decline as informative pixels are removed (shading = ± s.e.m). Humans perform better on EMIs across pixel visibility compared to Bubbles. **E.** Consistency between performance on masked stimuli and clean-image behavior for humans. EMI-based masks (blue) preserve higher behavioral consistency than Bubbles (red) across visibility levels, demonstrating that EMI perturbations recover the diagnostic features guiding human recognition. **F.** Cross-comparison of explanation consistency. Consistency between clean-image behavior and performance on EMIs is shown for humans (dark blue), models evaluated on the same EMIs (light blue), and human Bubbles data (red). Model behavior on EMIs parallels human performance, achieving Bubble-level consistency while requiring no behavioral trials.

## Cross-Species Validation: Testing MAPS in Monkeys and Neural Populations

As the closest available animal model of human object recognition[31], the macaque offers a powerful test bed for assessing whether MAPS-derived explanatory strategies reflect the underlying neural computations that support primate object recognition. We therefore conducted a two-stage validation linking model behavior, monkey performance, and neural population activity in inferior temporal (IT) cortex.

We first tested whether the model-derived strategies that best explained human behavior also generalized to two monkeys performing the same rapid match-to-sample object discrimination task (100 ms presentation, 100 ms delay; **Figure 7A**). Each monkey viewed the same clean image set used in humans, and we quantified image-level performance using the same behavioral metric (see *Methods: Behavioral Metrics*). As per MAPS, we determined that the most aligned ANN with monkey image-level behavior (**Figure 7B**) was Resnet-50 SSL[29,32]. We then compared the pooled empirical image-level behavioral accuracies of the monkeys on clean images (same 40 images as in the previous section) to the model's behavior on explanation-masked images (EMIs). As shown in **Figure 7C**, the best model–method pair (Resnet-SSL + Noise Tunnel Saliency) produced markedly stronger alignment with monkey behavior than the worst method (Deconvolution, one-sided paired t-test: $t(5) = 2.79, p = 0.019$). Notably, this correlation approached the empirical monkey–human correspondence (dashed line, $r(38) = 0.79, p < 0.001$), indicating that the explanatory structures recovered by MAPS capture visual strategies that are shared across species.

If these cross-species similarities reflect genuine computational alignment, the same best model–method pair should also predict activity patterns in the inferotemporal (IT) cortex—the neural substrate of object recognition in primates. To test this, we recorded multi-unit activity from IT populations while monkeys performed the same task and trained linear decoders to predict object identity from the neural responses (see *Methods: Deriving neuronal behavior on clean images*). We then correlated model behavior on EMIs with the performance of these neural decoders. As illustrated in **Figure 7D**, EMIs generated from the best attribution method preserved neural predictivity over a wide range of pixel budgets, whereas EMIs from the worst method led to a rapid decline in alignment (one-sided paired t-test: $t(5) = 2.91, p = 0.017$). The difference confirms that the model explanations identified as "best" by MAPS not only align with behavior but also with the representational computations implemented in the IT cortex.

Together, these results demonstrate that the explanatory strategies extracted by MAPS are not arbitrary artifacts of model training but reflect the representational logic shared across human and nonhuman primate vision. By linking model explanations to both behavior and neural population activity, MAPS bridges the gap between explainable AI and systems neuroscience—showing that the same explanatory principles can predict performance from pixels to neurons.

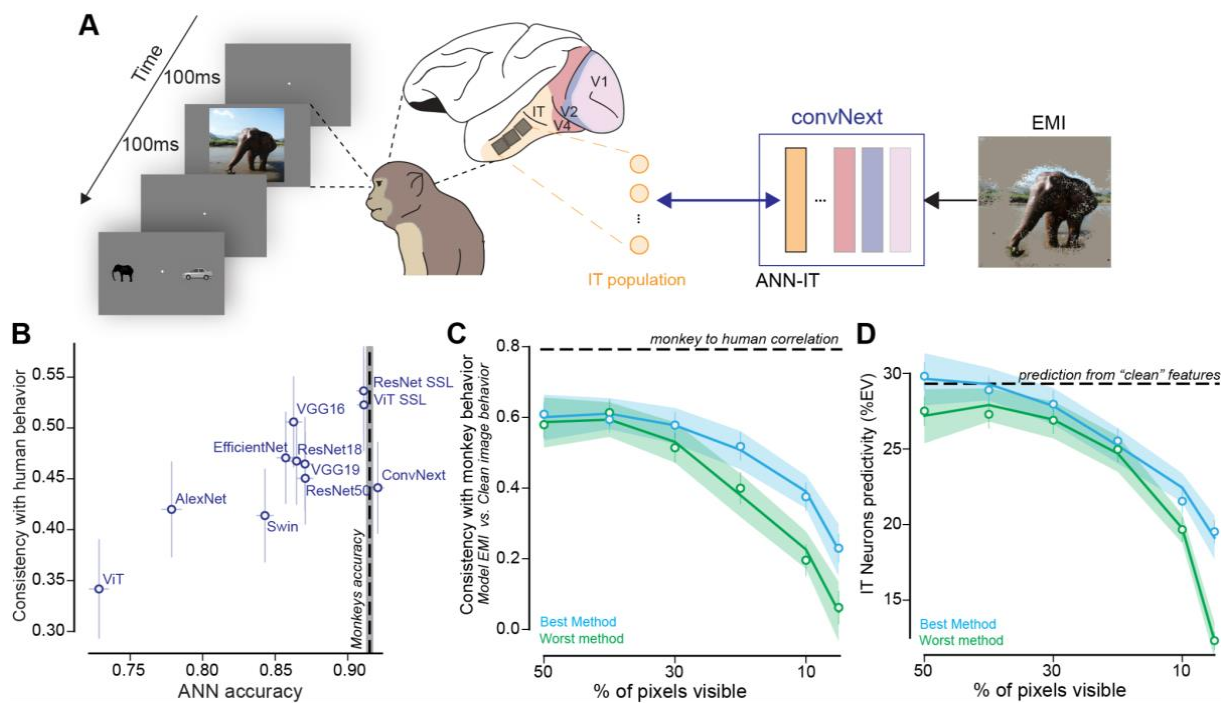

**Figure 7. Cross-species extension: MAPS links models, primate behavior, and neural populations. A.** Monkeys performed the same rapid object discrimination task (100 ms presentation, 100 ms delay) while fixating. Recordings were obtained from inferior temporal (IT) cortex populations. **B.** Relationship between model accuracy and consistency with monkey behavior. ANN accuracy correlates strongly with human recognition patterns (Pearson r = 0.80 , p = 0.003), suggesting that explanations derived from high-performing models may approximate monkey recognition strategies. The dashed horizontal line shows average monkey accuracy. We identify ResNet SSL as the accuracy-matched, most consistent ANN model. **C.** Correlation between monkey performance on clean images and model performance on EMIs (generated from the best vs. worst method) as a function of attribution percentile range. The best model–method pair (ConvNeXt + Noise Tunnel Saliency; blue) yields substantially higher correlations with monkey behavior than the worst method (Deconvolution; green). The dashed line indicates the empirical monkey–human behavioral correlation, showing that the best method approaches this cross-species benchmark. **B.** Predictivity of IT neural responses based on model activations from EMIs. The best model–method pair (ResNet SSL + Noise Tunnel Saliency; blue) yields significantly higher IT predictivity than the worst method (Deconvolution; green) across pixel-visibility levels. The dashed line denotes ResNet SSL's IT predictivity using clean-image features, illustrating that EMIs derived from the best explanations retain most of the neural correspondence observed under unperturbed conditions.

# Discussions

This study introduces MAPS (Masked Attribution-based Probing of Strategies)—a computational framework that transforms ANN model explanations into testable behavioral and neural predictions. At its core, MAPS asks whether the same image regions that drive a model's decision also drive primate perception and neural computation. By converting attribution maps from ANNs into explanation-masked images (EMIs), we were able to move beyond qualitative visualization and directly measure how explanatory structure influences recognition behavior across systems. Through a sequence of analyses that began with models, extended to humans, and culminated in monkeys and their IT cortex, MAPS revealed that the best ANN–explanation combinations reproduce not only human behavior but also the representational computations of the macaque ventral stream.

## Explanation variability and the need for a behavioral ground truth

The first step toward developing MAPS was to confront a basic but under-acknowledged fact: different explanation algorithms applied to the same model yield dramatically different attribution maps. Our systematic comparison across 11 architectures and 12 XAI methods (**Figure 2D**) showed that the variance across methods dwarfs the variance due to random initialization. This heterogeneity confirms that the choice of the explanation method can dominate the resulting inference about a model's decision making. MAPS provides a quantitative way to adjudicate among explanation methods by introducing a behavioral criterion. By first establishing a *ground-truth ranking* of explanation similarity directly from attribution maps and then showing that using EMIs one can recover this ranking through purely behavioral correlations, MAPS bridges introspective model analyses with externally measurable effects. The key insight is that behavioral perturbations can serve as a readout of explanatory fidelity. If two models respond similarly to identical EMI manipulations, their underlying attribution structures are functionally alike—even if their internal architectures differ.

## Recovering human visual strategies efficiently

Human object recognition depends on selective use of visual evidence[33], yet traditional psychophysical tools for identifying this selectivity—such as the Bubbles method[3] or classification images[34]—are exceptionally data-intensive. In our experiment, generating reliable Bubble maps for 40 test images required presenting ~10,000 randomly bubbled stimuli and collecting ~36,000 behavioral trials to estimate which set of pixels consistently contributed to recognition. By contrast, MAPS required no additional human trials. We simply need the baseline clean-image behavioral dataset (~4,000 trials), to serve as a validation reference for both MAPS and Bubbles (**Figure 6F**).

Explanation-masked images (EMIs) generated from the best ANN–XAI combination (ConvNeXt + Noise Tunnel Saliency) reproduced the hallmark "Bubble-like" behavioral pattern: an extended plateau of high recognition accuracy followed by a sharp decline (**Figure 6D**). This pattern indicates that the model-identified pixels align with the evidence humans actually use for recognition, demonstrating that attribution-based masking captures meaningful perceptual strategies.

Crucially, model behavior on EMIs predicted human image-by-image performance on clean images with accuracy comparable to the Bubbles approach (**Figure 6F**). This means that the tens of thousands of behavioral trials required to obtain Bubble-based importance maps can be eliminated entirely, with MAPS recovering equivalent human-aligned strategies through purely in silico manipulations. More broadly, MAPS generalizes to any visual decision task. Once a highly predictive ANN model and a validated attribution method are established for the desired behavior, EMIs can be generated and evaluated without new human data—shifting human testing from discovery to lightweight validation.

## Cross-species generalization and implications for systems neuroscience

Extending MAPS to the macaque system demonstrates that the explanatory strategies recovered from models capture not only human-like behavior but also core principles of primate visual computation. The macaque provides a critical bridge between behavior and neurophysiology: it performs the same object recognition tasks as humans[31] and affords direct access to the underlying neural circuitry in the IT cortex[35–37]. The observation that MAPS-derived best model–method combination aligns with both monkey behavior and IT population responses indicates that MAPS uncovers explanatory structures that are not species-specific artifacts but rather reflect a shared representational logic linking perceptual behavior to the cortical computations that support it.

This cross-species correspondence strengthens a primary claim of the framework—that the validity of an explanation extends beyond reproducing output patterns to capturing the causal features that constrain biological vision. The finding that the same explanation method which best predicts primate behavior also maximizes IT predictivity suggests that MAPS isolates a subset of features that the ventral stream computations deem behaviorally relevant. In doing so, MAPS transforms abstract attribution maps into empirically testable hypotheses about which visual features are represented and weighted by neuronal populations during recognition.

From a neuroscience standpoint, this approach offers a new way to interpret neural activity. Traditional model–brain comparisons—based on representational similarity[38,39] or accuracy[7,23,40,41]—quantify what is represented for behavior. MAPS provides a causal probe: by selectively removing pixels that a model identifies as explanatory, one can observe whether the corresponding behavioral or neural system degrades in parallel. This enables direct testing of whether the same information that governs model decisions also constrains biological vision.

Such perturbation-based comparisons open the door to addressing long-standing questions about hierarchical visual processing, and feedback processing[42]. If early visual areas are sensitive to low-level features while higher areas encode object-level abstractions, MAPS-derived EMIs can reveal which level of explanatory content best predicts neural degradation under visual masking. Because EMIs can be generated from any model layer, MAPS naturally complements layer-wise alignment approaches and can help identify the representational stage whose explanatory structure most closely mirrors that of the brain.

Importantly, MAPS does not assume that models and brains share identical architectures or feature bases. Instead, it evaluates whether the functional consequences of feature removal are consistent across systems—an emphasis that shifts the focus of computational neuroscience from architectural mimicry to behaviorally grounded functional alignment. By linking model explanations, behavior, and neural population responses within a single experimental framework, MAPS offers a unified and empirically testable pathway for bridging explainable AI and systems neuroscience.

## Implications for AI interpretability and benchmarking

From an artificial intelligence perspective, MAPS reframes the heterogeneity of interpretability method outcomes as a measurable, testable quantity rather than a visual artifact. By showing that certain attribution methods produce explanations that generalize across independently trained networks (**Figure 5C**), across species (**Figure 7C**), and across levels of representation (**Figure 7D**), MAPS provides an empirical basis for selecting among competing XAI techniques. This has several consequences. First, it identifies Noise Tunnel Saliency as a particularly robust method for recovering transferable visual strategies, offering a benchmark for future interpretability work. Second, it suggests a path toward standardized interpretability evaluation, where explanation quality is judged not by human intuition but by quantitative alignment with biological data. More broadly, MAPS contributes to a growing paradigm shift in AI: the move from accuracy optimization toward strategy alignment. In practical terms, this means training and evaluating models not only for performance but also for the degree to which their decision-relevant features mirror those used by humans[43]. Such alignment is essential for applications where interpretability and trustworthiness matter—ranging from clinical diagnostics to autonomous systems—because it ensures that models attend to causally meaningful evidence rather than spurious correlations.

## Methodological and conceptual advantages

MAPS offers a series of methodological and conceptual advances that move beyond descriptive explainability toward experimentally testable alignment between models and biological vision. First, it introduces objective behavioral grounding as the primary validation principle for explanations. Rather than judging attribution maps by visual inspection or by correlating them with subjective measures such as eye movements or manually labeled importance regions, MAPS evaluates explanation quality through their *quantifiable impact on behavior*. Each attribution map is transformed into a family of explanation-masked images (EMIs), and the resulting change in recognition performance—whether in a model, a human participant, or a monkey—serves as a direct, objective measure of how much the identified pixels contribute to the task. This approach defines explanation validity in measurable, falsifiable terms and anchors interpretability to observable behavioral outcomes. Second, MAPS achieves scalability by exploiting the model's own generative capacity to create stimuli. Traditional psychophysical approaches, such as the Bubbles method, require thousands of human trials per image to infer diagnostic features. MAPS replaces this manual process with model-guided EMI generation, allowing hundreds of targeted perturbations to be tested automatically per image. This design makes it possible to evaluate multiple attribution methods and image sets efficiently while requiring only a modest amount of human or animal behavioral data for validation. Third, MAPS introduces an element of causal testing into the analysis of visual explanations. Because EMIs selectively manipulate image regions that a model deems important, the resulting change in recognition performance constitutes an experimental intervention—a test of sufficiency and necessity for the features identified by the explanation. In this way, MAPS converts what is typically a correlational description of importance into an explicit measure of causal contribution, aligning model evaluation with the perturbation-based logic of neuroscience. Finally, MAPS enables cross-system comparability through a shared experimental stimulus space. The same EMI sets generated from a reference model can be presented to humans, monkeys, and other models, providing a direct way to compare behavioral and neural consequences under matched visual perturbations. This unified design allows researchers to determine whether biological and artificial systems depend on similar subsets of information for recognition, or whether their strategies diverge under identical constraints. Together, these properties establish MAPS as a principled, experimentally tractable framework for quantifying visual strategies.

## Limitations and future extensions

Despite its strengths, MAPS has several limitations that point the way to future work. First, it currently relies on a predefined library of attribution methods[44]. Although our analyses identify robust differences among them, newer gradient-free or concept-based approaches[45] may reveal additional explanatory dimensions. Future extensions[45] ould integrate these methods within the MAPS validation loop to continuously refine the behavioral ground truth for explanation quality. Second, the present implementation focuses on rapid object recognition, a canonical but static task. Natural vision, however, unfolds over time and integrates context[41], motion[46], and expectation[47]. Extending MAPS to videos and temporally evolving decisions— would allow examination of temporal visual strategies, such as how attention and memory shape evidence use. Third, while our cross-species results demonstrate strong correspondence between model EMIs and IT activity, causal validation remains an open frontier[48]. Combining MAPS with neural perturbations—for instance, inactivation or optogenetic disruption of regions corresponding to EMI-identified features—could reveal whether the same causal dependencies govern both artificial and biological systems. Finally, while our analyses focus on the ventral visual stream, the same logic could extend to other cognitive domains. Language models[49], for example, could generate *token-masked inputs* to test whether human comprehension patterns follow similar explanatory constraints. In motor control or decision-making, EMIs could be replaced by action-masked trajectories, probing whether AI agents rely on the same evidence or priors as humans. In each case, the core insight remains: the explanatory structure of a model can be tested through its behavioral consequences.

## Broader implications

By aligning explainable AI with psychophysics and systems neuroscience, MAPS unifies multiple research traditions into a single experimental framework that quantifies how selective information use shapes behavior and neural computation. Cognitive psychology has long shown that human recognition relies on sparse diagnostic cues[3], neuroscience reveals distributed[50] yet selective coding[37] in the IT cortex; and AI attribution methods visualize which inputs drive model outputs[14]. MAPS integrates these insights, showing that explanation is not an epistemic add-on but a biological constraint on how intelligent systems operate. By quantifying this constraint, MAPS tests whether models recognize objects for the same reasons primates do—shifting alignment from mimicking outputs to reproducing the rules of evidence selection. Through scalable, behaviorally grounded perturbations, MAPS connects models, humans, and neurons under a common empirical standard, revealing shared explanatory principles across species and systems. Extending this logic to dynamic vision[46], closed-loop neuroscience[51,52], and self-supervised learning[53] could refine our understanding of how representations emerge and are used, transforming explainability from description into an experimentally testable theory of representation that links pixels to neurons and algorithms to minds.

# Methods

## Visual Stimuli

We used images from the MS COCO imageset[54]. The images contained 10 object types (bear, elephant, face, apple, car, dog, chair, plane, bird, zebra). We split the images into training and testing sets, containing 1400 and 200 images, respectively, with equal representation of the 10 classes in each set. Images were processed at their native resolution and displayed in color.

## Data Collection

We carried out extensive behavioral data collection to validate our approach against established baselines. Participants completed the tasks on MTurk, an online crowdsourcing platform, for a payment of $15 CAD/hour. This experimental protocol involving human participants was approved by and in concordance with the guidelines of the York University Human Participants Review Subcommittee. Humans did not receive any additional training on the task. To manage the large number of experimental conditions, we tested a subset of 200-images test set, corresponding to 20 exemplars per object category. The reliability of the online MTurk platform has previously been validated by comparing results obtained from online and in-lab psychophysical experiments. In addition, we directly assessed the quality of our data by computing split-half reliability for each experimental condition (see *Trial split-half reliability estimation* below), and we ensured that the internal reliability of every dataset exceeded 0.8.

Each trial began with a 100-ms presentation of the sample image, followed by a 100-ms blank gray screen, and then a choice screen containing the target object and a distractor. Participants indicated their choice by mouse click or touch. To prevent repetition effects, each subject saw each image only once.

Behavioral data were collected under four experimental conditions:

- **Clean images baseline:** Fifty-eight participants each provided at least 108 responses per image, establishing object recognition performance.
- **EMI-Based experiment:** A total of 56 unique participants took part. Best and worst model–method pairs were evaluated with at least 36 responses per image on 40 images and 6 percentile value. This yields a total of *8,640* unique trials.
- **Object-only experiment:** Fourteen participants each provided at least 36 responses per image to assess recognition without contextual information.
- **Bubbles-based experiment:** Seven experiments (one per pixel-visibility level) were conducted with 20, 17, 21, 23, 21, 23, and 18 participants, respectively. Each participant provided at least 36 responses per image, for a total of 9,960 unique trials across pixel-visibility conditions.

## Behavioral Testing

### Non-human Primates

We have two adult male rhesus macaques (Macaca mulatta) as research subjects in our experiments. All data were collected, and animal procedures were performed, in accordance with the NIH guidelines, the Massachusetts Institute of Technology Committee on Animal Care, and the guidelines of the Canadian Council on Animal Care on the use of laboratory animals and were also approved by the York University Animal Care Committee. Some of the neural and behavioral data used in this study have been utilized in previous publications[9,5,64].

In the behavioral experiment, monkeys performed an object recognition task. Images were presented on an iPad (for details of home cage testing system, see[55]).During behavioral testing images were presented on an iPad. All images were shown at 8 deg of visual angle. Monkeys touched the screen to initiate a trial. Similar to the human participants, the trial started with the presentation of a sample image from the set of 1320 images for 100 ms. This was followed by a blank gray screen for 100 ms, after which the choice screen was shown displaying a target and a distractor image. The monkey was allowed to view the choice screen freely for up to 1500 ms and indicated its final choice by touching the selected image. Prior to testing, monkeys were trained to perform the delayed match to sample tasks on the same object categories (but with a different set of images, see[56] for details). Behavioral performance was quantified on an image-by-image basis, as the proportion of correct responses for each image across all distractors (see *Behavioral Metrics* below for more details).

**Behavioral Metrics**

We have used a one-vs-all image-level behavioral performance metric, **B.I$_1$**, similar to previous studies, to quantify the behavioral performance of monkeys and humans. This metric estimates the overall object discriminability of each image containing a specific target object from all other objects (pooling across all 9 possible distractor choices).

As mentioned above, for each trial of the task, a specific sample image was shown, and a binary choice task screen was presented. So, the data obtained from each trial is a correct (1) or incorrect (0) choice from the subjects. Each trial can be labeled with 2 unique identifiers – the unique sample image (one out of 40) and a unique task (one out of nine possible tasks given the image, e.g. bear vs. dog).

Hence, given an image of object 'i', and all nine distractor objects (j≠i) we computed the average performance per image (each element of the **B.I$_1$** vector) as the average of the percent correct across all the binary tasks done with that image as the sample image (where object 'i' was the target and all objects j≠i were the distractors respectively).

While the **B.I$_1$** vector provides an estimate of humans' image-by-image accuracy, the overall performance accuracy can be determined by taking an average of the **B.I$_1$** vector (across all images).

**Trial split-half reliability estimation**

To compute the reliability of the B.I$_1$ vector, we split the trials per image into two equal halves by resampling without substitution. The mean of the correlation of the two corresponding vectors (one from each split half), across 100 repetitions of the resampling was then used as the uncorrected reliability score (i.e., internal consistency), r. To correct for the usage of half the number of trials to estimate the reliabilities in comparison to the raw correlations, we used the Spearman Brown correction method on the uncorrected reliability score (r) as follows,

$$\text{Corrected reliability} = \frac{2 \times r}{1 + r}$$

**Noise ceiling estimation**

The noise ceiling is computed as the square root of the product of the trial split-half reliabilities of each variable used in the raw correlation. For instance, to estimate the noise ceiling (maximum correlation expected) for the comparison of one model vs. human pool B.I$_1$, the two relevant internal reliabilities are those estimated for B.I$_1$ of the model (set to 1 because the model is deterministic) and human pool, respectively.

## Analyses of ANN Models

**Model selection**

We selected a broad and diverse set of 11 artificial neural network (ANN) models spanning a wide range of architectures, training objectives, and inductive biases. These models include classic convolutional networks such as AlexNet[57], VGG16 and VGG19[30], and ResNet variants (ResNet50 and ResNet18)[29] as well as deep architectures like ConvNeXt[22] and EfficientNet[58]. We also included modern transformer-based architectures including Vision Transformers (ViT)[59] and Swin Transformers[60]. To capture the effects of training, we incorporated models trained under standard supervised learning, self-supervised learning (e.g., SimCLR[32]). All models were pre-trained on the ImageNet[61] object classification dataset and task (except for minor differences in training specifics as per the original publications), and we used the publicly available pre-trained weights. Details of training and models are available in https://github.com/vital-kolab/maps.

**Training Procedure on Images**

Images were standardized to RGB, resized to 224 × 224 pixels, and normalized using ImageNet channel means and standard deviations.

Training was performed on the final classification layer only, while all other layers remained frozen. Optimization was carried out with Adam optimizer and an initial learning rate of 0.001. Cross-entropy loss was used as the objective function. We used a training set (see *Visual Stimuli*) of 1400 images, split into batches of size 32, and training proceeded for 5 epochs.

**Generation of the explanation**

We computed attribution maps (also called saliency maps), which assign an importance value to each input pixel, indicating how strongly it contributed to the model's decision. Attribution methods typically propagate gradients or apply perturbations to measure how sensitive the output class score is to local changes in the input image. These maps therefore provide a spatial readout of the model's internal decision-making process.

For each trained model, we generated attribution maps on the 200 held-out test images (see *Visual Stimuli*) using gradient- and perturbation-based methods implemented in Captum (https://captum.ai/). The suite comprised Saliency[10] (vanilla gradients), Deconvolution[27], Input×Gradient[27], Guided Backpropagation[12], Integrated Gradient[13], GradientSHAP[27], Occlusion[11], Feature Ablation[62], Feature Permutation[28], and NoiseTunnel wrappers applied to gradient methods[64,65].

**Target - Reference pairs**

For each target model, we identified reference model whose behavior most closely matched that of the target on clean images. To do so, we computed the Spearman correlation between the image-level behavioral vectors (B.I$_1$) of the target model and those of all other models. This procedure was repeated across multiple cross-validated image splits (see *Cross-Validation* section) to ensure robustness, and the resulting correlation values were averaged across splits. The model yielding the highest mean correlation was selected as the reference for that target. The complete set of selected reference–target pairs is shown in **Figure S2A**.

The same procedure was applied using humans as the target, correlating human B.I1 with all models on clean images. In this case, the *ConvNeXt* model emerged as the best-matching reference to human behavior (see **Figure 2B**).

**Quantifying the variance between methods**

To assess the relative contributions of attribution method and model initialization to variability in explanations (**Figure 2D**), we computed pairwise L2 distances between attribution maps. For each of the 200 test images, attribution maps were first normalized (z-scored per map, with a small constant added for numerical stability) and then compared by computing the L2 norm of their difference. We carried out two complementary analyses. In the between-method condition, we compared attribution maps produced by different explanation techniques (e.g., Saliency vs. Guided Backpropagation) applied to the same model initialization; these distances reflect how strongly the choice of explanation algorithm alters the attribution pattern. In the between-initialization condition, we compared attribution maps produced by the same explanation method across different random initializations of the network; these distances reflect how much stochastic variation in training changes the explanations when the attribution method is held constant. For each image, we thus obtained two distributions of distances, one indexing method-driven variability and the other initialization-driven variability.

## MAPS Set up

**Ground truth (L2 similarity of explanations across model pairs)**

For each attribution method $m$ and each model pair $(M_{Reference}, M_{Target})$, we defined the ground truth as the mean L2-based similarity between their attribution maps over the 200 test images. Concretely, for image $i$, we computed a per-image similarity:

$$s_i^{(m)}(M_{Reference}, M_{Target}) = \frac{1}{||A_{Reference,i}^{(m)} - A_{Target,i}^{(m)}||_2 + \epsilon}$$

With a small constant $\epsilon = 10^{-8}$ to avoid division by zero. The ground-truth score for method $m$ and pair $(M_{Reference}, M_{Target})$ is then the average over images:

$$GT^{(m)}(M_{Reference}, M_{Target}) = \frac{1}{N}\sum_{i=1}^{N} s_i^{(m)}(M_{Reference}, M_{Target}), N = 200$$

and stacking $\{GT^{(m)}(M_{Reference}, M_{Target})\}_m$ yields a method-wise ground-truth vector for that model pair. This vector is the target against which we correlate the proxy (behavior-based) vectors in the final analysis.

**Generation of Explanation-Masked images (EMIs)**

To causally probe the role of explanatory features identified by each model, we generated Explanation-Masked images (EMIs) from the attribution maps of the reference models. For each attribution method and percentile threshold (percentile values $p$ : 1,2,3,5,10,15,20, 25,30,35,40,50), we first took the absolute value of the scores to capture the overall strength of contribution and then normalized each channel to the range [0,1]. A binary mask was then created by selecting the top-$p\%$ of pixels across channels, such that a pixel was considered explanatory if it exceeded the percentile threshold. This mask was used to construct positive EMIs, in which only the selected explanatory pixels were preserved, and all other regions were replaced with a uniform gray value. By design, an EMI with $p = 5$ retains only the top 5% most explanatory pixels, focusing the stimulus on the regions most strongly driving the model's decision. These manipulations were performed systematically for every test image, model, method, and percentile, providing a controlled way to examine how the removal of model-identified features influences behavioral outputs.

**Proxy based on behavioral similarity across models**

For each reference–target model pair, we asked whether the two models behave similarly when confronted with the same EMI stimuli. For every attribution method and percentile, we generated EMIs from the reference model's explanations and presented these identical stimuli to both models. Image-level behavior (model **B.I$_1$**, see also *Behavioral Metrics*) was summarized by an image-level score grounded in the match-to-sample paradigm: for each image, we estimate the model's probability to choose the target over the distractors. This yields, for each model, a vector of image-wise behavioral patterns. Behavioral similarity is then quantified, per attribution method, as the Spearman correlation between the reference and target B.I$_1$ vectors. Aggregating these correlations across methods produces a method-wise proxy profile for each percentile, which we compare against the corresponding ground-truth profile derived from L2 similarity of the two models' attribution maps.

**Comparing Proxy and Ground Truth**

To evaluate whether behavioral similarity serves as a reliable surrogate for explanation similarity, we directly compared the proxy rank-order with the ground-truth rank-order. For each reference–target model pair, attribution method, percentile, the proxy profile is given by the vector of Spearman correlations between the models' image-level behaviors across methods, while the ground-truth profile is the vector of L2-based similarities between their attribution maps across the same methods. We then quantified alignment by computing the Spearman correlation between the proxy and ground-truth vectors, thereby asking whether methods that produce more similar attribution maps between models are also those that yield stronger behavioral similarity. For each target model, we then reported the maximum alignment value over all percentiles as a validation of the proxy: it demonstrates that, for each reference, there exists at least one percentile value where the behavioral similarity profile closely tracks the distance between the models' explanation maps.

# Large-Scale Neural Recordings in the Inferior Temporal Cortex of Macaques

**Surgical Implants and Microelectrode Arrays**
We surgically implanted three 10×10 microelectrode arrays (Utah arrays, Blackrock Microsystems) per hemisphere in the inferior temporal (IT) cortex of each monkey under aseptic conditions. Each array contained 96 electrodes, excluding corner electrodes, with a length of 1.5 mm and a spacing of 400 μm between electrodes. We determined the placement of the arrays intraoperatively using the visible sulcus patterns for guidance. For monkeys receiving implants in both hemispheres, we initially implanted arrays in one hemisphere and recorded data for approximately one year before explanting and reimplanting new arrays in the opposite hemisphere.

**Electrophysiological Recordings**

During each experimental session, multiunit neural activity was recorded continuously at a sampling rate of 20 kHz using an Intan RHD Recording Controller (Intan Technologies, LLC). The raw voltage signals were bandpass filtered offline using a second-order elliptical filter (300 Hz to 6 kHz, 0.1 dB passband ripple, 50 dB stopband attenuation), before being thresholded to obtain the multiunit spike events. A multiunit spike event was defined as the threshold crossing when voltage (falling edge) deviated by more than three times the standard deviation of the raw voltage values. The implanted arrays sampled a range of regions along the posterior-to-anterior axis of the IT cortex. For all analyses, we treated each recording site as a random sample from the broader IT population without considering the precise spatial locations of the electrodes. We analyzed neural responses averaged between 70 ms and 170 ms after image onset.

**Eye Tracking and Calibration**

During recording sessions, we monitored eye movements using video eye tracking (SR Research EyeLink 1000). Using operant conditioning and water reward, our subjects were trained to fixate a central white dot (0.2°) within a square fixation window that ranged from ±2°. At the start of each behavioral session, monkeys performed an eye-tracking calibration task by making a saccade to a range of spatial targets and maintaining fixation for 500 ms. Calibration was repeated if drift was noticed over the course of the session.

Real-time eye-tracking was employed to ensure that eye jitter did not exceed ±2°, otherwise the trial was aborted, and data discarded. Stimulus display and reward control were managed using the MWorks Software (https://mworks.github.io).

## Validating MAPS

**Object-only stimulus generation**

To evaluate recognition performance in the absence of contextual information, we created object-only stimuli from the test images. For each image, objects were manually annotated by drawing their boundaries using a polygonal selection tool. Pixels inside the annotated region were preserved, while all remaining pixels were replaced with a uniform mid-gray background (RGB value = 0.5). This procedure yielded object-only versions of all selected test images, in which only the object was visible and all scene or contextual information was removed. These stimuli were then used in behavioral experiments to probe the extent to which recognition is driven by object features alone.

**Bubbles stimulus generation**

To benchmark MAPS against established psychophysical tools, we generated stimuli using the Bubbles method. This approach randomly reveals small, Gaussian-shaped regions of an image ("bubbles"), enabling inference of the diagnostic features used for recognition.

For each base image, we created 50 masked variants (trials). On each trial, 20 circular Gaussian bubbles were randomly positioned across the image. Each bubble had a Gaussian spread of 20 pixels (standard deviation) and revealed the underlying image with full transparency at its center (peak value = 1.0). The remainder of the image was blended with a uniform gray background set to an intensity of 0.5 (on a 0–1 scale).

Bubble masks were combined using a union rule, in which the visibility of a given pixel was computed as:

$1 - \prod_{k=0}^{20}(1 - b_k)$ where $b_k$ is the Gaussian contribution of the $k^{th}$ bubble at that pixel location.

This formulation ensures that overlapping bubbles increase the probability of revealing the underlying image, while all revealed regions remain bounded between 0 and 1 in transparency.

For each trial, we stored both the masked stimulus and the corresponding visibility mask, allowing later reverse-correlation analyses of behavior.

**Combination EMIs derived from Bubble performance**

To generate human-derived combination EMIs, we began with the set of fifty Bubble masks associated with each base image and the corresponding behavioral accuracy on those trials. Masks that yielded higher-than-average accuracy were weighted more strongly, whereas those linked to poorer performance contributed little or nothing. The weighted masks were summed and normalized to form a composite importance map, which was then blurred with a Gaussian kernel and passed through a logistic soft-knee nonlinearity to reduce noise and sharpen spatial structure. From this continuous map, we selected the most informative pixels, defined as the top X% of the distribution (X was varied from 5 to 50), and blended these regions with the original image while replacing the remainder with a uniform mid-gray background. Edges were softly tapered to avoid artificial boundaries. The resulting stimuli emphasize the regions of each image that human Bubble performance identified as most diagnostic for recognition, under a fixed pixel budget and with controlled visual appearance.

**Model-to-Neuron Prediction**

To assess how well model-derived representations could predict neural activity in the inferior temporal (IT) cortex, we performed a model-to-neuron mapping analysis. Model features were extracted from the layer previously identified as the best match to IT in the *Brain-Score* benchmark[21]. For each experimental condition, the *ResNet SSL* model was presented with EMIs generated using either the *Noise Tunnel Saliency* or *Deconvolution* attribution method. EMIs were generated at multiple pixel-visibility levels (i.e., varying the percentile of pixels retained from the attribution map), providing a graded manipulation of the visual evidence available to the model. This procedure allowed us to evaluate how model-to-neuron correspondence changes as a function of the visual information preserved in the image.

For comparison, neural data were obtained from monkeys performing a visual recognition task with the same set of clean (unaltered) images. To align the two domains, we performed linear mapping from model unit activations (model features) to neural firing rates recorded in the IT cortex. Specifically, for each neuron, we fit a linear regression model (ridge regression) that predicted the neuron's response to each clean image from the corresponding model feature vector. The mapping procedure was implemented using the open-source *Predictivity* toolbox (vital-kolab/reverse_pred), which provides standardized tools for fitting, cross-validating, and evaluating linear model–neuron mappings[66]. Model performance was assessed using percentage of explained variance (%EV), defined as the squared Pearson correlation between predicted and actual firing rates, normalized by the geometric mean of their split-half reliabilities (see Split-Half Reliability section), and scaled to percentage. We computed %EV for each neuron independently using the images from each pixel percentile derived from the Noise Tunnel Saliency and Deconvolution method. The model's overall IT predictivity score was defined as the mean %EV across the full population of neurons recorded.

## Statistical Analyses

All statistical analyses were performed in Python using scipy.stats. Relationships between continuous variables were assessed using Spearman correlation coefficients unless otherwise noted.

**Statistical Testing**

All statistical comparisons were guided by a consistent decision procedure based on the distributional properties and pairing of the data. First, we assessed whether each distribution was approximately normal using the Shapiro-Wilk test. If both groups satisfied the normality assumption, we applied parametric tests. For paired data (e.g., within-subject or within-model comparisons), we used a paired t-test (two-tailed; degrees of freedom = $n - 1$). For unpaired comparisons (e.g., between models), we used an independent samples t-test (degrees of freedom = $n_1 + n_2 - 2$).

If at least one distribution violated the normality assumption, we used non-parametric alternatives. Specifically, we applied the Wilcoxon signed-rank test for paired comparisons and the Wilcoxon rank-sum test (equivalent to the Mann–Whitney U test) for unpaired data. All tests were two-tailed unless stated otherwise. We report exact p-values and test statistics throughout.

**Decoders Cross-Validation**

A three-fold cross-validation scheme was used, dividing the dataset into training and testing subsets. The dataset comprised 200 images distributed equally across ten object categories (bear, elephant, person, car, dog, apple, chair, plane, bird and zebra). This design ensured a balanced representation of each category during training and testing.

## Cross-Validation

To ensure the robustness and generalizability of all reported results, we implemented a cross-validation scheme based on multiple random splits of the test set. First, we isolated from the set of 200 MS-COCO images, the 40 images used for psychophysics experiments. Then, we generated 20 independent two-way splits of this image set. Each split divided the 160 test images into two non-overlapping subsets of equal size. The first set of splits (20 folds, 80 images) was used for model and method selection. Within each fold, we computed behavioral correlations across models and attribution methods to identify (i) the best reference model for each target model (i.e., the model yielding the highest mean correlation across folds; see Target - Reference pair), and (ii) the best/worst-performing attribution methods (i.e., the method maximizing or minimizing the behavioral correlation across target models).

The second set of splits (20 folds, 80 images) was used exclusively for evaluation, ensuring an unbiased assessment of the selected references and methods. For each held-out split, we computed (i) the correlation between the ground truth and the proxy (see Comparing Proxy and Ground Truth), and (ii) the correlations of target behavior on clean images and target behavior on EMIs with the best and worst methods across percentiles (**Figure S3**). This two-stage approach—model/method selection on one set of splits and evaluation on an independent set—prevents overfitting and provides a cross-validated estimate of predictive performance.

# Supplementary Material

MAPS: Masked Attribution-based Probing of Strategies – A computational framework to align human and model explanations

**Authors**

Sabine Muzellec, Yousif Kashef Alghetaa, Simon Kornblith, and Kohitij Kar

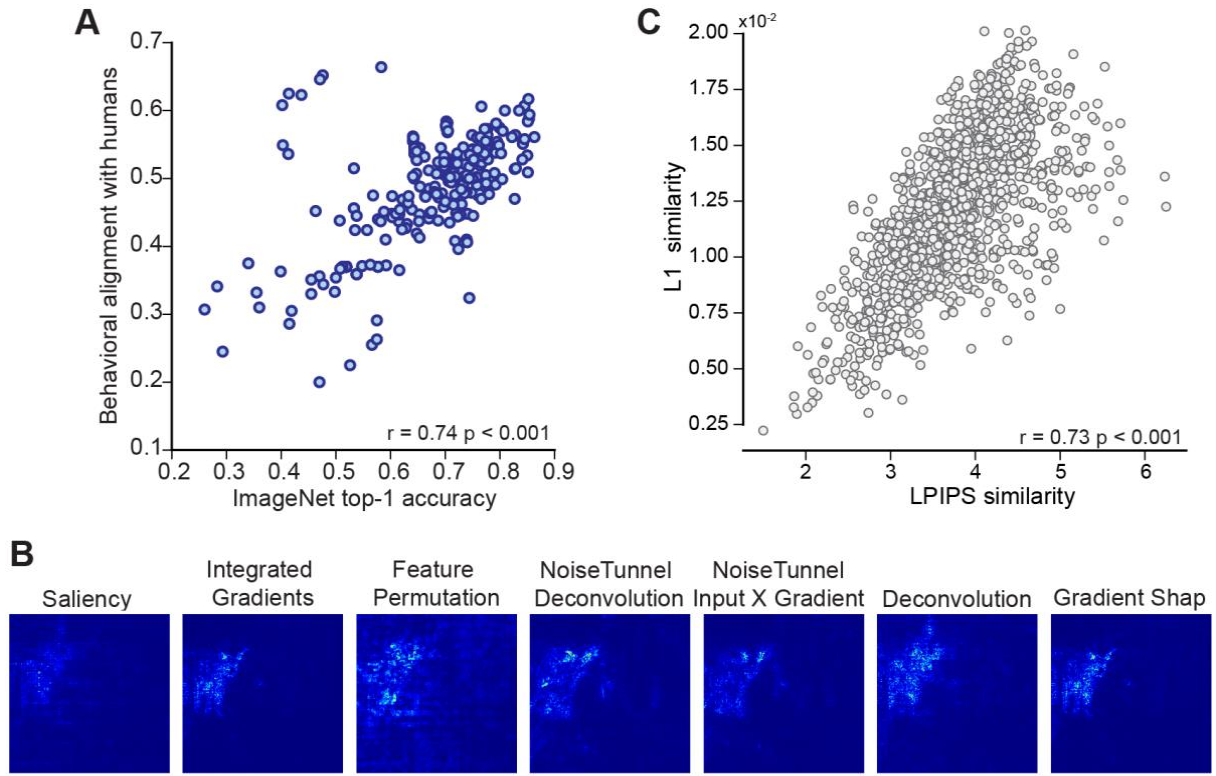

**Figure S1. A.** Relationship between model accuracy on ImageNet[61] and consistency with human behavior[7]. Data for the plot is adap[61]d from http://brain-score.org. ANN accuracy correlates strongly with human recognition patterns (humans: $r(261) = 074$, $p < 0.001$), even on image sets different from ours. **B.** Additional examples of attribution maps for the same image across the remaining XAI methods (Saliency, Integrated Gradients, Noise Tunnel Deconvolution, Noise Tunnel Input x Gradient, Gradient Shap and Feature Permutation). **C.** Correlations between alternative similarity metrics. Pairwise comparisons show that L1 and LPIPS distances are strongly correlated across images and model pairs, complementing **Figure 3B**.

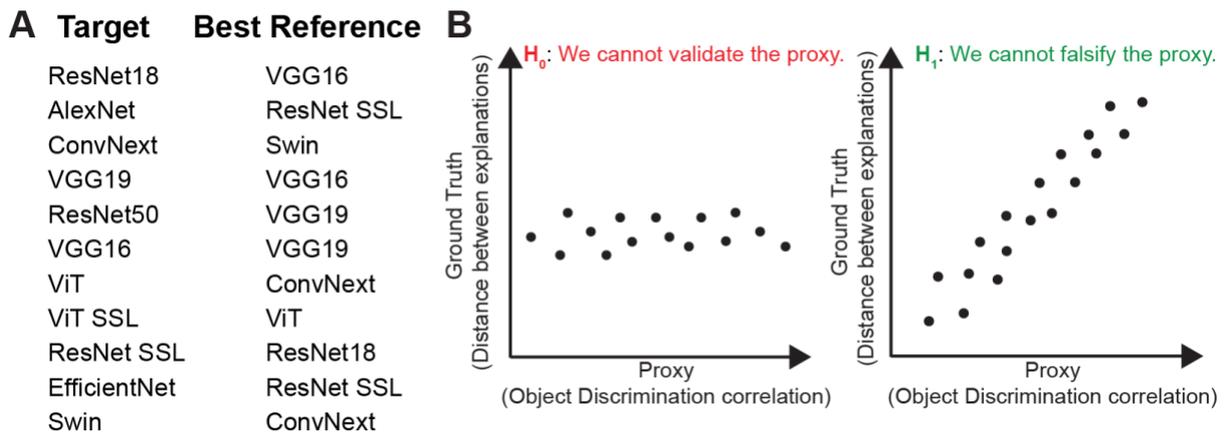

**Figure S2. A.** List of References for each Target model selected based on the behavioral correlation on clean images. **B.** Schematic of proxy validation logic. If proxy scores are unrelated to the ground-truth distances between explanations (left), the method fails. If proxy scores recover the ground-truth ranking (right), the proxy is supported.

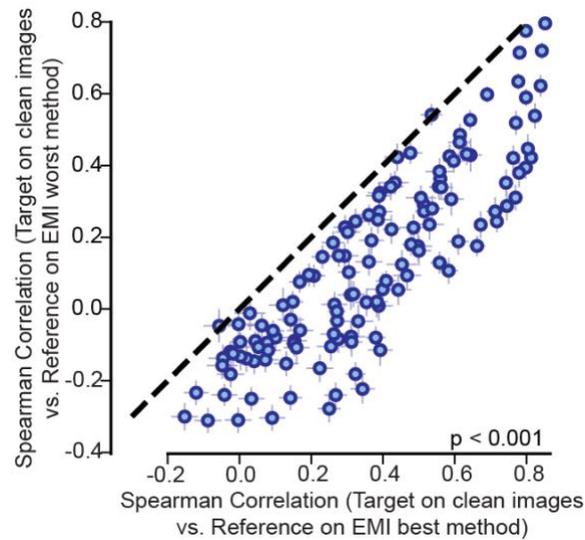

**Figure S3. A.** Comparison between the correlations obtained using the best and worst attribution methods. Each point represents a reference–target model pair and a percentile value, showing the correlation between their behaviors on clean images when the reference model's EMIs were generated with either the best (x-axis, Noise Tunnel Saliency) or worst (y-axis, Deconvolution) attribution method. The distribution shows a significant shift above the diagonal ($p < 0.001$), confirming that the best methods recover more meaningful behavioral relationships.

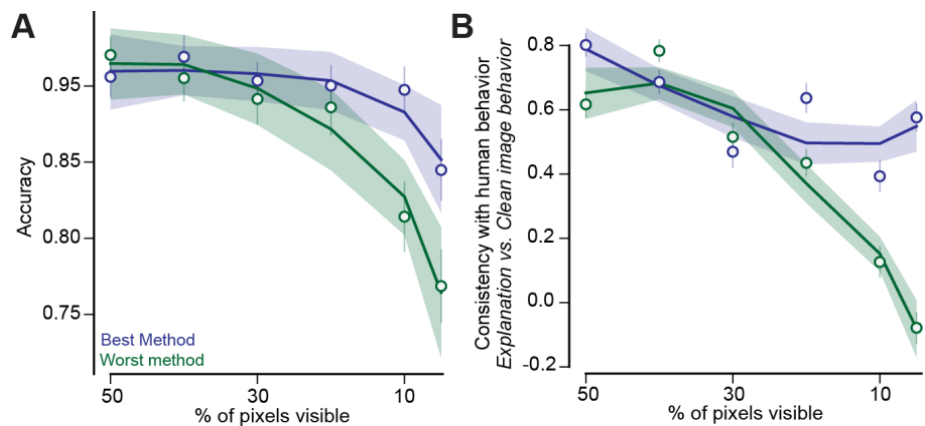

**Figure S4. A.** Human accuracy on EMI stimuli as a function of the percentage of visible pixels for the best (green, Noise Tunnel Saliency) and worst (blue, Deconvolution) attribution methods. As pixel visibility decreases, accuracy declines more steeply for the worst methods, indicating that EMIs derived from stronger attribution methods preserve more task-relevant information. **B.** Consistency between human behavior on EMIs vs. clean images as a function of visible pixel percentage. While consistency remains relatively stable for the best methods, it rapidly decreases for the worst methods, demonstrating that effective explanation-based perturbations maintain behavioral alignment with unperturbed image performance.

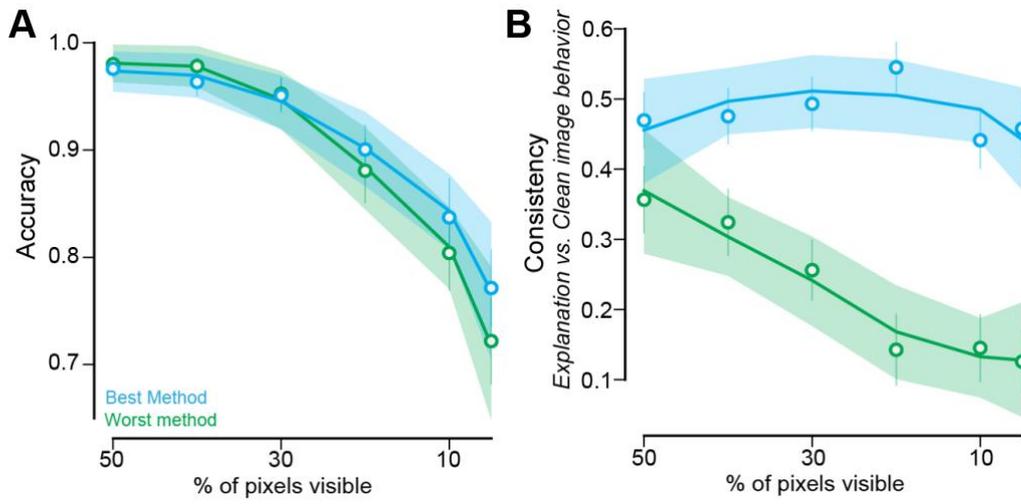

**Figure S5. A.** Model (ConvNeXt) accuracy on EMI stimuli as a function of the percentage of visible pixels for the best (green, Noise Tunnel Saliency) and worst (blue, Deconvolution) attribution methods. As pixel visibility decreases, accuracy declines more steeply for the worst methods, indicating that EMIs derived from stronger attribution methods preserve more task-relevant information. **B.** Consistency between human behavior on clean images vs. model behavior on EMIs as a function of visible pixel percentage. While consistency remains relatively stable for the best methods, it rapidly decreases for the worst methods, demonstrating that effective explanation-based perturbations maintain behavioral alignment with unperturbed image performance.